\documentclass[11pt]{article}
\pdfoutput=1
\newcommand{\PNrevA}[3]{#2}%{{#2}}% invisible, final
\newcommand{\PNrevAfn}[2]{#2}%{{#2}}% invisible, final

\newcommand{\PNrevB}[3]{#2}%{{#2}}% invisible, final
\newcommand{\PNrevBfn}[2]{#2}%{{#2}}% invisible, final

\newcommand{\PNrevC}[3]{{#2}}% invisible, final
\usepackage{xcolor,graphicx,amsmath}
\usepackage{eucal}  % see http://www.ctan.org/tex-archive/fonts/amsfonts/doc/euscript.pdf
\usepackage{pifont,amsmath,bm,amsfonts,dcolumn,cancel,upgreek} %
\usepackage{amssymb}  % for some special characters
\usepackage{chemarr}  % chem rxn arrows
\usepackage[pdftex,%pageanchor=false,% needs to be true for index to work
plainpages=false,pdfpagelabels,%
colorlinks=true,citecolor=grayzero,urlcolor=grayzero,linkcolor=grayzero,bookmarks=true,bookmarksopen=true,
bookmarksnumbered=true,pdfstartpage=2,pdfdisplaydoctitle=true,
naturalnames=true]{hyperref}
\hypersetup{pdftitle={Proofreading Nelson et al.},pdfauthor={P. Nelson}}
\definecolor{grayzero}{gray}{0} %also called black

\def\cut#1{}\def\pnonly#1{}%{{\tiny\color{red}\bf #1}}

\DeclareSymbolFont{MathSL}{OT1}{cmr}{m}{sl}
\DeclareSymbolFontAlphabet{\mathsl}{MathSL}
\newcommand{\ex}[1]{{\mathrm e}^{#1}}                 % exponential

\newcommand{\ee}[1]{\cdot10^{#1}}
\newcommand{\eem}[1]{\cdot10^{-#1}}
\newcommand{\dd}{{\mathrm d}}
\newcommand{\tot}{_{\mathrm{ tot}}}

\newcommand{\lessim}{\protect\raisebox{-0.75ex}[-1.5ex]{$\;\stackrel{<}{\sim}\;$}}
\newcommand{\inv}{^{\raise.15ex\hbox{${\scriptscriptstyle -}$}\kern-.05em 1}}
\newcommand{\exv}[1]{{\langle #1\rangle}}
\newcommand{\ssr}[1]{_{\rm #1}} % "subscript roman"

\newcommand{\bn}{\thinspace}         %separate big numbers (not comma)

\def\BDsynVAL{{0.15}} % synthesis rate of the birth-death process that matches Golding data, 1/min
\def\burstsizVAL{5}

\DeclareMathOperator{\variance}{var}
\newcommand{\ssref}[2][Sect.]{#1\thinspace\ref{ss:#2}} %cite subsection (\label{ss:n.m.k})
\newcommand{\aref}[2][Appendix]{#1\thinspace\ref{a:#2}} %cite subsection (\label{ss:n.m.k})
\newcommand{\fref}[2][Fig.]{{#1\thinspace\ref{f:#2}}}    %cite figure (from \ifigfull{f:xx}..)
 \newcommand{\frefpanel}[3][Fig.]{{#1\thinspace\ref{f:#2}}\capitembare{#3}}    %cite figure (from \ifigure{f:xx}..)
\newcommand{\eref}[2][Eq.]{#1\thinspace\ref{e:#2}}      %cite equation (from \label{e:n:xx}
\let\ereffarcomma\eref
\newcommand{\arttitle}[1]{#1}
\newcommand{\capitem}[1]{(\textsf{#1})\enspace }\newcommand{\capitembare}[1]{#1}%!NO: {\textsf{#1}}
\newcommand{\capitemrefer}[1]{(#1)}
\newcommand{\capelement}[1]{\emph{#1}}%
\newcommand{\FIindex}[1]{#1}
\newcommand{\Nindex}[1]{}
\newcommand{\FNindex}[1]{}
\newcommand{\instructor}[1]{}%{{\tiny #1}}
\def\<#1>{{\small\tt #1}}
\newcommand{\artfrom}[1]{(#1)}

\newcommand{\Npoiswords}{Poisson distribution\Nindex{Poisson!distribution}}

\newcommand{\Ndrvbare}{\ensuremath{n}} % generic discrete r.v.

\newcommand{\NSrate}{\Npprate\ssr s} % synthesis
\newcommand{\Nratecon}{{k}}\newcommand{\NrateconX}[1]{{\Nratecon_{#1}}} %!
\newcommand{\NDratecon}{\NrateconX{\Nzilch}}
\newcommand{\Nzilch}{\mathrm{d}}%\mathrm{\char'34}}           % for network diagrams
\newcommand{\Npprate}{k}  % mean rate of poisson process 
\let\NX\Ndrv
\let\Nx\Ndrv
\newcommand{\NXrm}{{{\mathrm{\Ndrvbare}}}}

\newcommand{\sunit}{\ensuremath{\mathrm s}}
\newcommand{\minunit}{\ensuremath{\mathrm{min}}}
\newcommand{\kcalunit}{\ensuremath{\mathrm{kcal}}}
\newcommand{\moleunit}{\ensuremath{\mathrm{mol}}} 
\newcommand{\uMunit}{\ensuremath{\mu{\scriptstyle\mathrm M}}}

\def\kpc{k'\ssr c}
\def\kc{k\ssr c}
\def\lpc{\ell'\ssr c}
\def\lc{\ell\ssr c}

\def\kpw{k'\ssr w}
\def\kw{k\ssr w}
\def\lpw{\ell'\ssr w}
\def\lw{\ell\ssr w}

\def\mpc{m'\ssr c}
\def\mc{m\ssr c}
\def\mpw{m'\ssr w}
\def\mw{m\ssr w}
\def\kadd{k_{\rm add}}
\def\kaddc{k_{\rm add,c}}
\def\kaddw{k_{\rm add,w}}

\def\GTP{\ensuremath{\mathrm{GTP}}}
\def\GDP{\ensuremath{\mathrm{GDP}}}
\def\C{\ensuremath{\mathrm{C}}}
\def\CGTP{\text{C$\cdot$GTP}}
\def\CGDP{\text{C$\cdot$GDP}}
\def\Pi{\ensuremath{\mathrm P_\mathrm i}}
\def\R{\ensuremath{\mathrm R}}
\def\W{\ensuremath{\mathrm W}}
\def\WGTP{\text{{W}$\cdot${GTP}}}
\def\WGDP{\text{{W}$\cdot${GDP}}}

\newcommand{\Nprob}{{{{\cal P}}}} % probability

\begin{document}
		\title{Stochastic Simulation to Visualize Gene Expression and Error Correction in Living Cells\\%{\small [[or\\Stochastic Simulation Exercises that Improve Understanding of Fundamental Cellular Processes]]{}}
		}
	\author{Kevin Y.\ Chen\\
	Department of Chemistry, University of Cambridge,\\ Lensfield Road, Cambridge CB2 1EW, UK\\
	Daniel M. Zuckerman\\
	Dept of Biomedical Engineering,
	Oregon Health \emph{\&} Science Univ,\\
	Portland, OR 97239,\\ and 
	Philip C.\ Nelson\\
	Department of Physics and Astronomy, University of Pennsylvania, \\Philadelphia PA 19104}
	\date{\today}\maketitle
\begin{abstract}
Stochastic simulation can make the molecular processes of cellular control more vivid than the traditional differential-equation approach by generating typical system histories instead of just statistical measures such as the mean and variance of a population. Simple simulations are now easy for students to construct from scratch, that is, without recourse to black-box packages. In some cases, their results can also be compared directly to single-molecule experimental data.
After introducing the stochastic simulation algorithm, this article gives two case studies, involving gene expression and error correction, respectively. Code samples and resulting animations showing results are given in the online supplement.
\end{abstract}
\noindent 
%\cut
{%
%Need to cite \cite{Wallin:2013cj}.\\
\color{red}
% Before final: Eliminate notation516 and put all needed defs in the main file. RevTex. Outline fonts. Grayscale all black objects. Enforce biblio style. RevTeX. http://web.mit.edu/rhprice/www/Contributors/manFormat.html
% Please package the manuscript and figure files into a single .zip archive for submission.
% Clean up code (DONE 1,2) and repost to github; update links as needed. Also upload the npz files that were used to generate the videos. \\
% "Graphs of formulas or data should ordinarily be plotted inside a frame, with tick marks around all sides and clear labels on both axes."
% Name each figure file in the format "AuthorNameFig01.eps". Please package the manuscript and figure files into a single .zip archive for submission.
% Most figures in AJP are printed at a width of one text column, about 3.4 inches (8.6 cm). When necessary, a wide figure can be printed across both columns.
% DONE "Number tables using Roman numerals"
% DONE ``Capitalize the first word of each text label, but for a multi-word label, capitalize only the first word and any proper nouns (as in an ordinary sentence).''\\
% DONE ``AJP  discourages the use of color to convey essential information.  the caption of a color figure, and any other text that refers to it, should not refer in any essential way to the colors used in the figure. When appropriate, you may include a parenthetical "(color online)" in the figure caption.''
}

\section{Introduction}
Physical processes unfold over time. Our minds grasp physical mechanisms largely via narrative. So it is not surprising that some of the most vivid physics demonstrations also play out over time. Simulations of physics that unfold over time are similarly powerful; interactive simulations are better; and simulations created by the student can be best of all. This view is gaining ground in introductory courses
\cite{Bchab15a}, but the benefits of animated simulation extend farther than this. Here we wish to show that the  behavior of strongly nonequilibrium statistical systems can be illustrated via stochastic simulations that are simple enough to serve as undergraduate projects. Recently developed, free, open-source programming resources sidestep the laborious coding chores that were once required for such work. 
In particular, we believe that the error-correction mechanism known as kinetic proofreading can be more clearly understood when a student views its linear temporal sequence, as opposed to solving deterministic rate equations. Coding this and other simple processes opens the door for the student to study other systems, including those too complex for the rate-equation approach to yield insight.

\section{Double-well hopping}
\subsection{The phenomenon}
We tell students that a simple chemical reaction, for example isomerization of a macromolecule, can be regarded as a barrier-passing process. A micrometer-size bead in a double optical trap serves as a mesoscopic model system with this character \cite{Simon:1992cr}, and it is well worthwhile for students to watch it undergo a few dozen sharp transitions in between episodes of Brownian motion near its two stable positions (see supplementary video \ref{sv1}).
A simple model for this behavior states that the hopping transitions occur at random times drawn from an exponential distribution. That is, many rapid transitions are interspersed with a few long pauses\cut{To put this a different way, "activated processes" which involve crossing barriers significantly greater than $k_B T$ tend to be rare, not slow.}.

\subsection{Simulation: Waiting times drawn from an exponential distribution\label{ss:wtde}}
With this physical motivation, students can explore how to generate simulated waiting times of the sort just described. Any computer math system has a pseudorandom number generator that generates floating-point numbers uniformly distributed between 0 and 1. Many students are surprised (and some are intrigued), to learn that applying a nonlinear function to samples from a random variable yields samples with a different distribution, and in particular that $y=-\tau\ln x$ is exponentially distributed, with mean $\tau$, if $x$ is uniform on (0,1] \cite{Bnels15a}\cut{This construction can be nicely illustrated with a graphic that shows its generality, especially the cdf approach.}.

Starting from that insight, it takes just one line of code to generate a list of simulated waiting times for transitions in a symmetric double well; finding the cumulative sums of that list gives the actual transition times (see supplementary computer code \ref{sc1}). 

The freely accessible VPython programming system (or its Web-based version Glowscript) makes it very easy to create an animation of an object whose spatial position is supplied as a function of time \cite{Evpyt17a}. The only challenging part is to pass from a list of irregularly-spaced transition times to particle positions at each of many (regularly-spaced) video frames (see supplementary computer code \ref{sc1}). The payoff is immediate: Visually, the simulated hopping has a very similar character to the actual Brownian hopping of a bead in a double trap (see supplementary video \ref{sv2}).

\subsection{Upgrade to 1D random walks}
It may be of interest to make a small modification of the code: Instead of hopping between two wells (reversing direction on every step), consider one-dimensional diffusion on a symmetric many-well potential, for example, one of the form $U(x)=\sin(x)$. In such a potential, for each transition the system must also decide whether to increase or decrease a ``position'' coordinate. The resulting random walk will display the same long-time scaling behavior as any unbiased 1D walk, but with trajectories that undergo hops at random times, not periodic steps as in the simplest realization \cite{Bnels15a}. 

\section{Birth-death process}
We can now generalize from situations with essentially only one kind of transition (or two symmetric kinds), to the more interesting case where several inequivalent choices are possible, and where the relevant probabilities depend on the current state. This general situation can describe a chemical reaction that requires, and depletes, molecules of some substrate.

Most science students know that living cells synthesize each of their messenger RNAs ({mRNAs}) from a single copy (or a small fixed number) of the corresponding gene. Even a constitutive (unregulated) gene must wait for the transcription apparatus to arrive, bind, and begin transcription \cite{Balbe09a}. We consider a situation in which that apparatus is in short enough supply that this waiting is the primary determinant for the initiation of transcription. Once a mRNA transcript has formed, it has a limited lifetime until it is degraded by other cellular machinery. We assume that this process, too, relies on chance encounters with degradation enzymes. Moreover, each of many species of mRNA must all share the attentions of a limited number of degradation enzymes, so each mRNA copy has a a fixed probability per unit time to be removed from the system.

The physical hypotheses in the preceding paragraph amount to a model called the ``birth-death process,'' which has many other applications in physics and elsewhere. As in the 1D walk, we characterize the system's state by an integer, in this case the population of the  mRNA of interest. Synthesis is a transition that increases this number, with a fixed probability per unit time $k_\mathrm{s}$ (called the ``mean rate'' of synthesis). Degradation is a transition that decreases it, with a probability per unit time that is the current population $n$ times another constant $k_\mathrm{d}$ (the ``rate constant'' for degradation)\cut{I suggest including the governing differential equation.  That will help make these statements concrete for students who should be comfortable with dif eqns.}. 

\subsection{Simulation\label{ss:bdp.s}}
D.~Gillespie extended and popularized a simple but powerful method, the ``stochastic simulation algorithm,'' for simulating systems of this sort \cite{Gillespie:1976p1686}. In the case just described, the algorithm repeatedly executes the following steps (see supplementary computer code \ref{sc2}):\begin{itemize}
\item Determine the probability per time $k\tot$ for \emph{any} of the allowed transitions to occur by summing all the mean rates. In a birth-death process, we have $k_\mathrm{tot}=k_\mathrm{s}+nk_\mathrm{d}$. 
\item Draw a waiting time from the exponential distribution with mean given by the reciprocal of $k\tot$, via the method in \ssref{wtde}.
\item Determine which of the allowed processes happens at that transition time. In the birth-death process, we make a Bernoulli trial with probability $p=k_{\mathrm s}/k_\mathrm{tot}$ to increase population $n$ by one, and $1-p$ to decrease it.
\item Update $n$ and repeat.\end{itemize}
The beauty of this algorithm, besides its correctness \cite{Gillespie:1977p1466}, is that no computation is wasted on time steps at which nothing happened: By definition, there is a state transition at \emph{every} chosen time.

\subsection{Convergence to the continuous, deterministic approximation\label{ss:ccda}}
Students will probably find it reasonable that, when $n$ is sufficiently large, we may neglect its discrete character. Students who have been exposed to probability ideas may also find it reasonable that in this case, the relative fluctuations of $n$ from one realization to the next will be small, and so $n$ effectively behaves as a continuous, deterministic variable, subject to the differential equation $\dd n/\dd t=-k_\mathrm dn+k_\mathrm s$. That equation predicts exponential relaxation from an initial value $n_0$ to the steady value $n_*=k_\mathrm{s}/k_\mathrm{d}$ with e-folding time $k_\mathrm{d}$:
\begin{equation}n(t)=n_*+(n_0-n_*)\ex{-k_\mathrm{d}t}.
	\label{e:bdcts}\end{equation}
The simulation bears out this expectation (\frefpanel{f1}{a,b}).
\begin{figure}\begin{center}\includegraphics{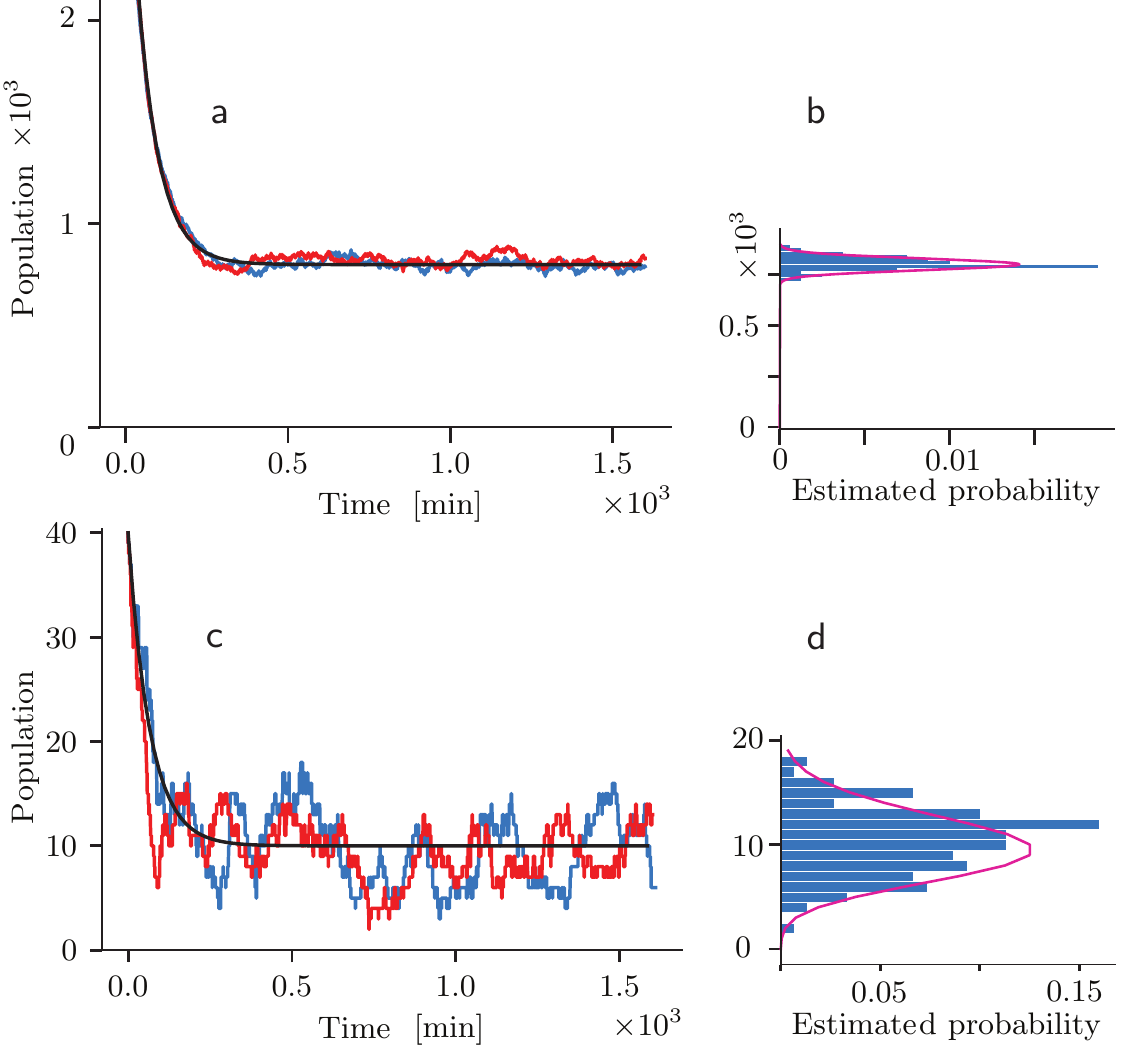}\end{center}
\caption{\label{f:f1}\small
\arttitle{Behavior of a birth-death process.}
\capitem aThe \capelement{bumpy traces} show two examples of simulated time series with $k_\mathrm{s}=12\,\minunit\inv$, $k_\mathrm{d}=0.015\,\minunit\inv$, and hence $n_*=800$. The initial population was $n_0=3200$.
The \capelement{smooth trace} shows the exponential relaxation predicted by the continuous, deterministic approximation (\eref{bdcts}).
\capitem bAfter the system comes to steady state, there is a tight distribution of $\NX$ values across 100 runs of the simulation (\capelement{bars})\cut{Bars are hard to see.  I suggest reducing vertical scale.  Also, on both axes, label multiple tick marks to avoid ambiguity of values.}. The \capelement{curve} shows the \Npoiswords{} with mean $n_*$
for comparison.
\capitem{c,d}The same but with $k_\mathrm{s}=0.15\,\minunit\inv$ and $n_0=40$. Although individual instances (runs of the simulation) deviate strongly from the continuous, deterministic approximation, nevertheless the sample mean of the population $\NX(t)$ over 150 runs does follow that prediction\cut{ (\capelement{green trace} in c)}. The distribution of steady-state fluctuations is again Poisson (\capelement{curve} in d).
(See also \cite{Bnels15a}.)
\instructor{Made by \<transcrip2rxn.py, viewTransc.py>.}
}\end{figure}

Actually, mRNA populations in living cells are often \emph{not} large. Nevertheless, although individual realizations of $n(t)$ may differ significantly, the \emph{ensemble average} of many such trajectories does follow the prediction of the continuous/deterministic idealization (\fref{f1}c). Within individual cells, there will be significant deviation around that mean behavior (\fref{f1}c again). In particular, the ``steady'' state will have fluctuations of $n$ that follow a Poisson distribution (\frefpanel{f1}{d}). That key result is more memorable for students when they discover it empirically in a simulation than it would be if they just watched the instructor prove it with abstract mathematics (by solving a master equation \cite{Bnels15a}).

State fluctuations of the sort just mentioned may suffice to pop a more complex system out of one ``steady'' state and into a very different one. Indeed, even the simplest living cells do make sudden, random state transitions of this sort. Such unpredictable behavior, not seen in the differential-equation approach, seems to be useful to bacteria, implementing a population-level ``bet hedging'' strategy \cite{Choi:2008p2537,Eldar:2010p9554,Lidstrom:2010gj}.
% Lidstrom ME, Konopka MC: The role of physiological heterogeneity in microbial population behavior. Nat Chem Biol 2010, 6:705-712.

A real bacterium is not simply a beaker of reagents. Bacteria periodically divide, partitioning a randomly chosen subset of each mRNA species into each daughter cell. That extra level of realism is hard to introduce into an analytical model, but straightforward in a simulation. The results are similar to the ones just described, with a larger effective value of the clearance rate constant \cite{Bnels15a}.

\subsection{Upgrade to cover bursting processes\label{ss:ucbp}}
Bacteria are supposedly simple organisms. The birth-death process is simple, too, and it fits with the cartoons we see in textbooks. So it is interesting to follow the recent discovery that the model makes quantitative predictions for mRNA production that were experimentally disproven \cite{Golding:2005p1478,Golding:2008p1486,Bnels15a}. 

For example, recent advances in single-molecule imaging permit the direct measurement of $n(t)$ in individual cells, and disproved the model's prediction that the distribution of $n$ in the ``steady'' state should be Poisson (\fref{f2}{c}). Researchers found, however, that a simple modification of the birth-death model could accommodate these and other discrepant data. The required extension amounts to assuming that mRNA transcripts are generated in \emph{bursts,}  that the bursts themselves are initiated with a fixed probability per unit time, and that once initiated, a burst is also terminated with a fixed probability per unit time. Although this ``bursting model'' has two additional parameters compared to the original birth-death model, nevertheless it was overconstrained by the experimental data, so its success was a nontrivial test \cite{Golding:2008p1486,Bnels15a}. Remarkably, detailed biochemical mechanisms for this behavior were found only some years after its indirect inference \cite{Chong:2014p15102,Sevier:2016ew,Sevier:2018ek,klindz}, an important lesson for students to appreciate.
\begin{figure}\begin{center}\includegraphics{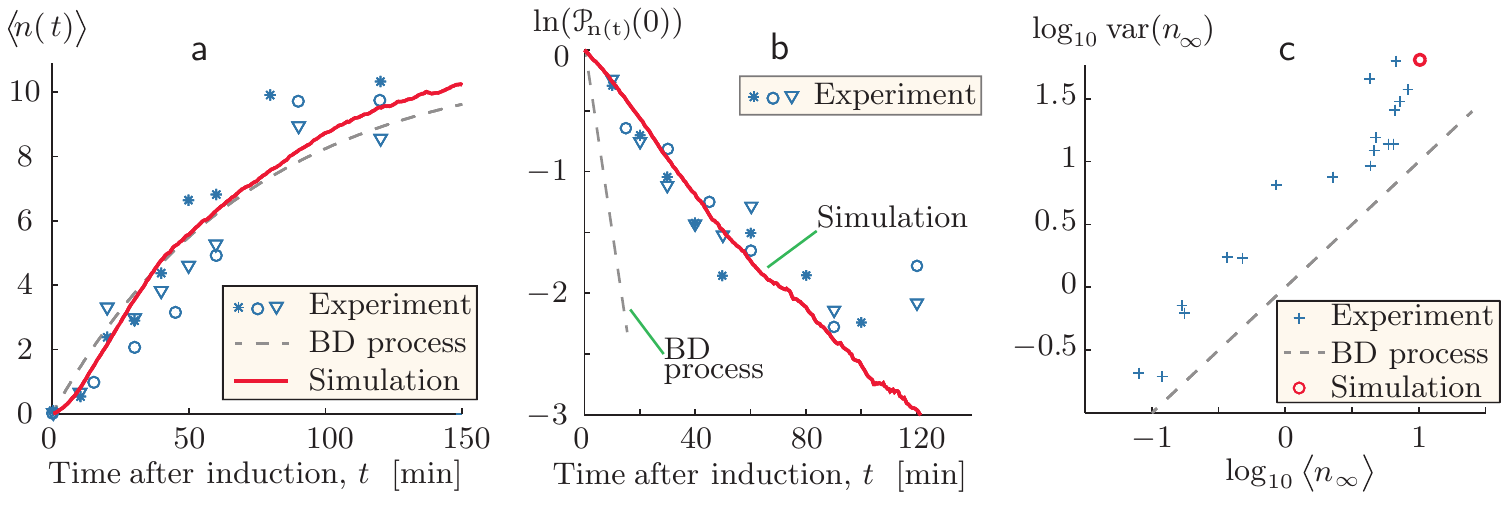}\end{center}
\caption{\label{f:f2}\small
\arttitle{Indirect evidence for transcriptional bursting.}
\capitem{a}\capelement{Symbols:} Time course of the number of {mRNA}\FNindex{RNA!messenger (mRNA)} transcripts in a cell, $\Nx(t)$, averaged over 50 or more cells in each of three separate, identical experiments. Data from the three trials are shown with three different symbols.
All of the cells were induced\FNindex{induction} to begin gene expression at  time zero. 
The \capelement{dashed curve} shows a fit of the birth-death (BD) process (\ereffarcomma{bdcts}) to data, determining the apparent synthesis rate $\NSrate\approx\BDsynVAL/\minunit$ and clearance rate constant $\NDratecon\approx0.014/\minunit$. The \capelement{solid curve} shows the corresponding result from a computer simulation\FNindex{simulation!random process (Gillespie algorithm)!bursting model} of the \FIindex{bursting} model discussed in \ssref{ucbp}  (see also \cite{Bnels15a}).
\capitem{b}Semilog plot of the fraction of observed cells that have zero copies of {mRNA} versus elapsed time. \capelement{Symbols} show data from the same experiments as in \capitemrefer{a}. \capelement{Dashed line:} The \FIindex{birth-death process} predicts that initially $\Nprob_{\NXrm\rm(t)}(0)$ falls with time as  $\exp(-\NSrate t)$, where $\NSrate$ has the value found by fitting the data in \capitemrefer a. (Degradation was negligible in this experiment.)
The experimental data instead yield initial slope $-0.028/\minunit$. \capelement{Solid line:} Computer simulation of the bursting model.
\capitem{c}Experiments were performed at each of many different levels of gene induction. For each level, a \capelement{cross} shows the variance of the late-time mRNA population $n_\infty$ versus its sample mean. This log-log plot of the data shows that they fall roughly on a line of slope $1$, indicating that the \FIindex{Fano factor} $(\variance\Nx)/\exv{\Nx}$ is roughly a constant. The simple \FIindex{birth-death process} predicts that this constant is equal to 1 (\capelement{dashed line}),  because mean equals variance for any Poisson distribution, but the data instead give the value $\approx\burstsizVAL$. The \capelement{circle} shows  the result of the same simulation shown in \capitemrefer{a,b}.
\artfrom{Figure adapted from \cite{Bnels15a}; experimental data from \protect\cite{Golding:2005p1478}\pnonly{Actually taken from \protect\cite{Golding:2008p1486}}\cut{; see \dref{bursting}}.}\instructor{Made by \texttt{bursting/mRNAsimMain.m} et al.}
}\end{figure}

	\section{Kinetic proofreading}\subsection{Word model\label{ss:wmkp}}
\instructor{``Already in 1957, before the advent of molecular biology, Linus Pauling postulated that this intrinsic (thermodynamic) limitation to the selection of similar amino acids would give rise to very large amino acid substitution errors in intracellular proteins [Pauling L: The probability of errors in the process of synthesis of protein molecules. Festschrift Arthur Stoll. Birkh\"ause Verlag; 1957:597--602.]'' -- Johansson etal\\
	``The free
energy differences due to base pair mismatches between the
mRNA codon and the tRNA anticodon are too small to provide
the observed high accuracy of tRNA selection (Ogle and Ramakrishnan,
2005; Xia et al., 1998), even if kinetic proofreading
(Hopfield, 1974; Ninio, 1975) is taken into account'' -- \cite{Savir:2013ez}}%
Ask a student, ``What is the big secret of life?'' and the answer will probably be ``DNA,'' or perhaps ``evolution by natural selection.'' Indeed, DNA's high, but not perfect, degree of stability underlies life's ability to replicate with occasional random modifications. But it is less well appreciated that the \emph{stability} of a {molecule} of DNA does not guarantee the \emph{accuracy} of its {replication} and {transcription.} There is another big secret here, just as essential to life as the well known ones. In fact, a wide range of molecular recognition events must have extremely high accuracy for cells and their organisms to function. Think of our immune cells, which must ignore the vast majority of antigens they encounter (from ``self''), yet reliably attack a tiny subpopulation of foreign antigens differing only slightly from the self.

Translation of mRNA into proteins is an emblematic example of such a puzzle. It is true that artificial machines now exist that can read the sequence of mRNA. Then another artificial machine can take the resulting sequence of base triplets, decode it, and synthesize a corresponding polymer of amino acids (a polypeptide), which in some cases will then fold into a functional protein without further help. But the cells in our bodies, and even bacteria, do these jobs reliably \emph{without}
those huge and expensive machines, despite the incessant nanoscale thermal motion! 

Merely intoning that a wonderful molecular machine called the ribosome accomplishes this feat doesn't get us over the fundamental problem: At each step in translation, the triplet codon at the ribosome's active site fits one of the many available transfer RNA (tRNA) species \emph{somewhat} better than it fits the other 19 options. 
But the binding energy difference, which quantifies ``somewhat better,'' only amounts to two or three hydrogen bonds. This translates into a fraction of time spent bound to the wrong tRNAs that is about $1/100$ times as great as the corresponding quantity for the correct amino acid \cite{Bphil08a}. If the fraction of incorrect amino acids incorporated into a polypeptide chain were that high, then \emph{every} protein copy longer than a few hundred amino acids would be defective!

In fact, the error rate of amino acid incorporation is more like $10^{-4}$. The fact that this figure is so much smaller than the one seemingly demanded by thermodynamics remained puzzling for decades. After all, the ribosome is rather complicated, but it is still a nanoscale machine. \emph{Which} of its features could confer this vast improvement in accuracy?

J.~Hopfield and J.~Ninio proposed an elegant physical mechanism, based on a  known but seemingly pointless feature of the ribosome \cite{Hopfield:1974p15095,Ninio:1975vv,Bphil08a}. To explore it, we begin by paraphrasing a metaphor due to U.~Alon \cite{Balon06a}. Imagine that you run an art museum and wish to find a mechanism that picks out Picasso lovers from among all your museum's visitors. You could open a door from the main hallway into a room with a Picasso painting. Visitors would wander in at random, but those who do not love Picasso would not remain as long as those who do. Thus, the concentration of Picasso lovers in the room would arrive at a steady value (with fluctuations, of course) that is enriched for the desired subpopulation.

To improve the enrichment factor further, you could hire an employee who occasionally closes the door to the main hallway, stopping the dilution of your enriched group by random visitors. Then open a new exit doorway onto an empty corridor. Some of the trapped visitors will gratefully escape, but die-hard Picasso lovers will still remain, leading to a second level of enrichment. After an appropriate time has elapsed, you can then reward everyone still in the room with, say, tickets to visit the Picasso museum in Paris.

The original authors realized that in the ribosome, the initial, reversible binding of a tRNA was followed by a transformation analogous to closing the door in the preceding metaphor. This transformation involved hydrolysis of a GTP (guanosine triphosphate) molecule complexed with the tRNA, and hence it was nearly \emph{irreversible,} due to the highly nonequilibrium concentration of GTP, compared to the hydrolysis products GDP and \Pi{} (inorganic phosphate).  Such hydrolysis reactions were well known to supply the free energy needed to drive otherwise unfavorable reactions in cells, but here their role is more subtle. 

Hopfield and Ninio were aware that after the hydrolysis, incorporation of the amino acid was delayed and could still be preempted by unbinding of the tRNA complex. The existence of this pathway had previously seemed wasteful: An energy-rich GTP had been ``spent'' without anything ``useful'' (protein synthesis) being done\cut{and indeed the hydrolysis actually created a delay and opportunity for reversing the step that had just been achieved!}. On the contrary, however, the authors argued that this second step implemented the mechanism in the art museum metaphor, giving the ribosome an independent second chance to dismiss a wrong tRNA that accidentally stayed bound long enough to progress to this stage. After all, spending some GTPs may be a modest price to pay compared to creating and then having to recycle an entire defective protein.

Hopfield coined the name ``kinetic proofreading'' for this mechanism, but we will refer to it as the ``classic Hopfield--Ninio'' (HN) mechanism because the original term is somewhat misleading. In chemical reaction contexts, a ``kinetic'' mechanism generally implies bias toward a product with lower activation barrier, even if it is less stable than another product with higher barrier. This preference is most pronounced at high, far-from-equilibrium catalytic rates \cite{Sartori:2013fv}. In contrast, the classic HN proofreading model involves two sequential thermodynamic (quasiequilibrium) discriminations\cut{, with the second acting on the biased output of the first}. Moreover, these discriminations take place prior to reading even the very next codon, in contrast to editorial proofreading, which generally happens after an entire manuscript is written. (Our choice of term also distinguishes the classic scheme from later models that are sometimes also called ``kinetic proofreading.'')

The qualitative word-model given earlier in this section may seem promising. But the corresponding kinetic equations make for difficult reading and understanding. Better intuition could emerge from a presentation that stays closer to the concrete ideas of discrete actors randomly arriving, binding, unbinding, and so on, visibly implementing the ideas behind the ``museum'' metaphor. The following sections will argue that stochastic simulation can realize that goal.
In a nutshell,
\begin{quote}\textsl{An effectively irreversible step, or at least a step far from equilibrium, gives rise to enhanced accuracy. The free energy of GTP hydrolysis is the price paid for this accuracy.}\end{quote}

It would also be valuable to confirm a key result of the analytic approach, which predicts that the enhancement of accuracy depends on GTP, GDP, and \Pi{} being held far from chemical equilibrium, so that the hydrolysis step is nearly irreversible (the ``door shuts tightly'' in the museum metaphor). In fact, the model predicts \emph{no enhancement} of accuracy when this chemical driving force is low  \cite{Hopfield:1974p15095}. Far from equilibrium, however, the predicted error fraction can be as low as the \emph{square} of the equilibrium value (or even a higher power if multiple rounds of sequential testing are employed). \pnonly{CLINCH}

\subsection{A single ribosome in a bath of precursors\label{ss:srbp}}
This section's goal is to formulate the word-model of \ssref{wmkp} in the context of mRNA translation, then set up a stochastic simulation (see also \cite{Ezuck17a}). Later sections will show how students can explore the expectations raised at the end of the preceding section.

\begin{figure}\hbox to 6in{\includegraphics{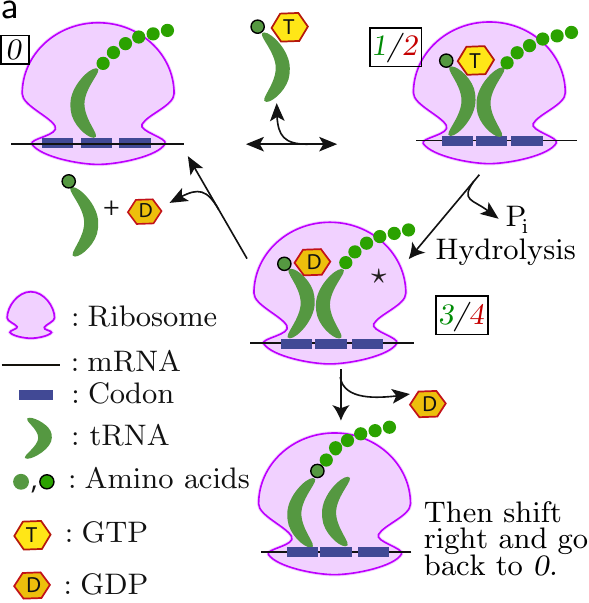}
\qquad\includegraphics{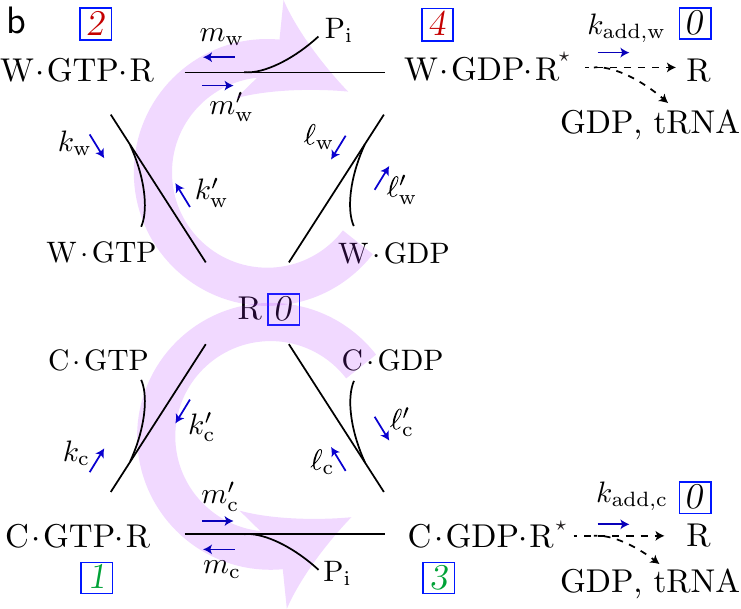}}
\caption{\label{f:f4}\small\arttitle{Two representations of the classic Hopfield--Ninio mechanism.}
\capitem aTraditional cartoon expressing the catalytic cycle of the ribosome (after \cite{Banerjee:2017en}).
\capitem bCorresponding kinetic diagram.
The large pale arrows indicate the net circulation in each cycle under cellular conditions, where \GTP{} is held out of equilibrium with \GDP{} and \Pi. The symbol R denotes a ribosome complexed with mRNA\PNrevBfn{ with first binding site either empty or bound to a GTP complex}{}; R$^*$ is the corresponding complex ``activated'' by GTP hydrolysis. At far right, R indicates the ribosome with  one additional amino acid added to the nascent polypeptide chain. The classic Hopfield--Ninio proofreading model assumes that unbinding rates $\kc$ and $\lc$ are smaller than their mismatched counterparts $\kw$ and $\lw$, but that other constants are all equal for correct and wrong tRNA.
}\end{figure}
We will assume that a single ribosome is complexed with a single mRNA and has arrived at a particular codon. This complex sits in an infinite bath containing several free, dissolved species at fixed concentrations (\fref{f4}a): \begin{itemize}
\item \C{} denotes \underline correct tRNA (that is, the species that matches the codon currently being read), loaded with the corresponding amino acid. We will neglect the possibility of a tRNA being incorrectly loaded; accurate loading is the concern of another proofreading mechanism that we are not  studying now \cite{Hopfield:1976p14958,Yamane:1977p14957}.
\item \W{} is similar to \C, but refers to the \underline wrong tRNA for the codon under study.
\item Some reactions form complexes of tRNA with guanosine phosphates: \CGTP, \CGDP, \WGTP, and \WGDP.  (For simplicity, we suppress any mention of elongation factors, one of which, ``EF-Tu,'' is also included in these complexes but is only implicit in the classic HN mechanism.)
\end{itemize}

\fref{f4}b denotes the ribosome-mRNA complex by \R. In state \textsl0, this complex is not bound to any tRNA. (More precisely, no tRNA is bound at the ``A'' site of the ribosome; a previously bound tRNA, together with the nascent polypeptide chain, is bound at another site (\fref{f4}a), which we do not explicitly note.) Surrounding this state, \fref{f4}b shows four other states $\mathsl1$--$\mathsl4$ in which the ribosome is bound to the complexes introduced earlier. The upper part of the figure describes wrong tRNA binding and possible incorporation; the lower part corresponds to the  correct tRNA. Horizontal arrows at the top and bottom denote hydrolysis of GTP, which is coupled to a transformation of the ribosome into an activated state, $\R^*$. 

Although any chemical reaction is fundamentally reversible, under cellular conditions the concentration ratio  $[\Pi][\CGDP]/[\CGTP]$ is far below the equilibrium value, so that the reactions in \fref{f4}b are predominantly in the direction shown by the pale arrows.
% the rate constant for $\C\cdot\GDP$ to bind to \R{} and form $\C\cdot\GDP\cdot\R^*$ is much smaller than for $\C\cdot\GTP$ to bind to \R{} and form $\C\cdot\GTP\cdot\R$. 
This was one of the conditions in Hopfield's original proposal. (\ssref{rtdf} will explore relaxing it.)

Again, we are assuming that a \emph{single} ribosome bounces around this state diagram in the presence of fixed concentrations of feedstocks either imposed in vitro by the experimenter or supplied by a cellular milieu. There are two ways to ``exit the museum exhibit by the second door'': After hydrolysis, the ribosome can reject its tRNA-GDP complex with probability per unit time $\ell$. Or, with probability per unit time $\kadd $ it can {add} its amino acid to the nascent polypeptide, translocate the tRNA to the second binding site, and 
eject any tRNA already bound there. Either way, the main binding site becomes vacant and, for the purposes of this state diagram, the ribosome returns to state \textsl0.

Supplementary computer code \ref{sc3} implements a Gillespie simulation on the five states of the ribosome (\fref{f4}b).
\pnonly{CLINCH}

\subsection{Visualization of the simulation results\label{ss:vsr}}
To keep the project modular, we constructed a simulation code that writes its state trajectory to a file.  A second code then reads that file and creates a visual output. The first of these codes operates similarly  to \ssref{bdp.s}, but with a four-way choice of what transition to make after each waiting interval.
The second code can be almost as simple as the one described in \ssref{wtde}. However, students with more time (perhaps in a capstone project) can make a more informative display with a reasonable additional effort, as follows.

The supplementary videos
not only show the state that is current at the end of each video frame; they also animate the pending arrivals of new complexes that are about to bind and the departures of old ones that have unbound without incorporation. By this means, the videos give a rough sense of the ``narrative'' in the trajectory being shown. These improvements are not difficult to add once the basic code is working. Alternatively, students can construct the basic version, then be shown these videos.

% \subsection{Stretching of short waiting times\label{ss:sswt}}
The exponential distribution of waiting times implies that there will be episodes with several events happening rapidly, interspersed with long pauses. For this reason, it is useful to view the simulation in two ways: Once with a shorter time step that resolves most individual events but covers only a limited time interval (supplementary video \ref{sv3}), and then with a coarser time step to see the entire synthesis trajectory (supplementary video \ref{sv4}). 

\begin{figure}\begin{center}
	\includegraphics{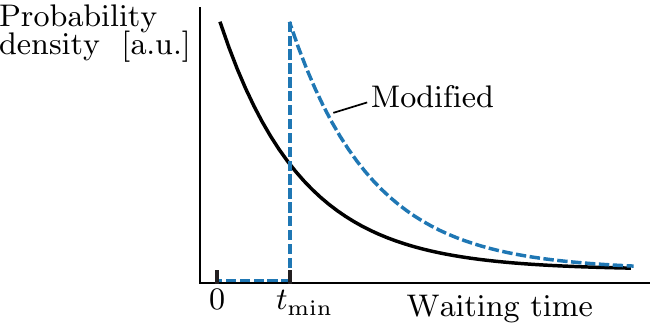}
	\end{center}
\caption{\small\arttitle{Modified waiting times.} \textit{Solid line:} Exponential distribution of waiting times. \emph{Dashed line:} Shifted exponential distribution obtained by adding the constant $t_\mathrm{min}$ to each sample.
\label{f:alterwait}}\end{figure}
We also found it useful  (solely for visualization purposes) to alter the distribution of waiting times in a simple way that relieves visual congestion without, we think, too much damage to the realism of the simulation. Our modification, shown in the supplementary videos, was to add a small fixed delay, for example one half of one video frame, to every transition waiting time (\fref{alterwait}).

\subsection{Classic Hopfield--Ninio model\label{ss:ckp}}
Following Hopfield, we initially assume that the rate constant for incorporation, $\kadd $, is the same regardless of whether the tRNA is correct or incorrect. We also   suppose that the binding rates $\kpc=\kpw$ and $\lpc=\lpw$ also have this property; for example, all of them may be diffusion-limited \cite{Bphil08a}. Only the \emph{unbinding} rates differ in the classic  HN model:
$$\kw=\phi_{-1}\kc\text{ and }\lw=\phi_3\lc.$$
Here $\phi_{-1}=\phi_3\approx100$ is the preference factor for unbinding the wrong tRNA (relative to the correct one).
Again following Hopfield, we will also take the hydrolysis rate constants to be equal:   $m'\ssr w=m'\ssr c$ (and $m\ssr w=m\ssr c$). 
\begin{table}\small\begin{tabular}{l|l|l||l|l|l}\textsl{Description}&\textsl{Symbol}&\textsl{Name in code}&\textsl{Classic HN}&\textsl{Realistic model}&\textsl{Equilibrium}%\ \  [s$\inv$]
	\\\hline
	binding GTP complex, $\mathrm s\inv\vphantom{1^{\strut}}$&$\kpc$&\texttt{kc\_{}on}&$40$&40&
	40\\
	unbinding GTP complex, $\mathrm s\inv$&$\kc$&\texttt{kc\_{}off}&$0.50$&0.5&
	0.5\\
	binding GDP complex, $\mathrm s\inv$&$\lpc$&\texttt{lc\_{}on}&\PNrevAfn{$2.36\eem{12}$}{0.001}&\PNrevAfn{$1.18\eem{10}$}{0.001}&
	\PNrevAfn{13.04}{0.26}\\
	unbinding GDP complex, $\mathrm s\inv$&$\lc$&\texttt{lc\_{}off}&$0.085$&0.085&
	0.085\\
	hydrolysis and P$_\mathrm i$ release, $\mathrm s\inv$&$\mpc$&\texttt{mhc}&$0.01$&25&
	\PNrevAfn{25}{0.01}\\
	condensation/P$_\mathrm i$ binding, $\mathrm s\inv$&$\mc$&\texttt{msc}&\PNrevAfn{$2.36\eem{12}$}{0.001}&\PNrevAfn{$1.18\eem{10}$}{0.001}&
	\PNrevAfn{13.04}{0.26}\\
\hline
	binding GTP $\vphantom{1^{\strut}}$&$\kpw=\phi_{1}\kpc$&\texttt{kw\_{}on}&$\phi_{1}=1$&$0.68$&
	1\\
	unbinding GTP &$\kw=\phi_{-1}\kc$&\texttt{kw\_{}off}&$\phi_{-1}=5$&94&
	5\\
	binding GDP &$\lpw=\phi_{-3}\lpc$&\texttt{lw\_{}on}&$\phi_{-3}=1$&0.0027\pnonly{$^\star$}&
	1\\
	unbinding GDP &$\lw=\phi_{3}\lc$&\texttt{lw\_{}off}&$\phi_{3}=5\pnonly{^\dagger}$&7.9&
	5\\
	hydrolysis&$\mpw=\phi_2\mpc$&\texttt{mhw}&$\phi_2=1$&0.048&
	1\\
	condensation&$\mw=\phi_{-2}\mc$&\texttt{msw}&$\phi_{-2}=1$&1&
	1\\
\hline
	incorporation$\pnonly{^\#}\vphantom{1^{\strut}}$, $\mathrm s\inv$&$\kaddc $&\texttt{kaddC}&%, but see \S\ref{ss:sswt}
	$0.01$&4.14&
	0.01\\
	incorporation&$\kaddw =\phi_\mathrm{add}\kaddc$&\texttt{kaddW}&$\phi_\mathrm{add}=1$ 	&$0.017$&
	1\\
% \hline\hline
\hline\end{tabular}
\caption{%Table I: 
\small Illustrative values for the rates shown in \fref{f4}b. The third column gives variable names used in supplementary computer code \ref{sc3}. See the Appendix for discussion of the numerical values. The fifth column follows \cite{Banerjee:2017en,ISI:000398884800036}, who refer to our $\phi_i$ as $f_i$.  The last column uses rates from the HN model, but with hydrolysis and incorporation modified to satisfy equilibrium. 
\hfil\break
% Above the double line, the fourth column gives estimated numerical values from \cite{Ezuck17a}; however, some alternative values have been chosen for more convenient viewing (\ssref{vsr}). Entries below the double line are computed from entries above it by \eref{KP.constr2}. In the last column, the rate constants have been multiplied by appropriate concentrations to yield apparent first-order rates in inverse secIonds.
\pnonly{Footnotes: $\star$:~Banerjee et al.\ chose $\phi_{-2}=1$, then got $\phi_{-3}$ by consistency% (their eqn. 2)
: $\phi_1\phi_2\phi_3=\phi_{-1}\phi_{-2}\phi_{-3}$. 
$\dagger$:~Hopfield chose $\phi_3=\phi_{-1}$. Here it's required by the consistency condition. 
$\ddag$:~$\mc$ and $\lpc$ were chosen to be 0.001 to ensure irreversibility of the GTP hydrolysis step \cite{ISI:000398884800036}.
$\#$:~\textbf{\color{red}Here Phil disagrees with Banerjee and used the rate they quote for the irreversible next-to-last step.}}}
\end{table}
To visualize wrong incorporations within a reasonable time frame, we raised the probability of incorrect choices: The preference ratios $\phi_{-1}$ and $\phi_3$ were lowered from their realistic value of 100 to just 5. Other values in column 4 of Table I were loosely inspired by rate constants estimated from experimental data  in a simplified form  (see Appendix \ref{a:1}). (\ssref{mrm} will follow the experimental values more closely.) \PNrevC{These effective rate constants are either zeroth order (unbinding and hydrolysis), or else
pseudo-zeroth order (binding and condensation), with rates appropriate for the concentrations of reactants present in the experiment.}{These effective rate constants are either a constant probability per unit time (unbinding and hydrolysis) or else a probability per unit time with the substrate concentration already lumped in (binding and condensation). The values we chose were appropriate for the concentrations of reactants present in the experiment.}{*} \PNrevB{Some of the rates were modified to satisfy conditions for the classic Hopfield--Ninio kinetic proofreading model. For example, $\mpc$ and $\kaddc$ were lowered from 25$\,\sunit^{-1}$ and 4.14$\,\sunit^{-1}$, respectively, to 0.01$\sunit^{-1}$  (see supplementary section X.X for a full discussion of the Hopfield model rate constants).}{}{*} 

Supplementary videos \ref{sv3}--\ref{sv4} show the resulting behavior. 
Perhaps the most important impression we get from viewing these animations is that \emph{the cell is a busy place.} The riot of activity, the constant binding events that end with no ``progress'' (and often not even GTP hydrolysis), are hallmarks of chemical dynamics that are hard to appreciate in textbook discussions, yet vividly apparent in the simulation. 
This is especially apparent in supplementary video \ref{sv4}, which shows a typical run of 25 amino acid incorporations. Because there are many unproductive binding and unbinding events in the simulation, not every event is shown in detail in video \ref{sv4}. However, focusing on the GDP-tRNA rejections shows that more correct tRNAs than incorrect tRNAs make it past GTP hydrolysis, and that the few incorrect tRNAs that do make it past are quickly rejected in the second proofreading step. In the instance shown, only one incorrect amino acid was incorporated out of 25 incorporations, much lower than the 1/5 error rate expected from single step equilibrium binding. Supplementary video \ref{sv3} provides a more detailed look at this process. 
The videos also show clearly the jerky, nonuniform progress of synthesis, with some amino acid incorporations happening after much longer delays than others. That feature is by now well documented by single-molecule experiments. 

A typical run created a chain of 100 amino acids, of which 6 were wrong. This error rate of $\approx(6/100)=0.06$ is \PNrevB{similar to the value $1/(\phi_3)^2 = 1/5^2 = 0.04$}{far smaller than the naive expectation of $1/\phi_3=0.20$}{*}. This is the essence of the classic HN mechanism; we see it taking shape in the animation, as many wrong tRNA complexes bind but are rejected, either prior to or after GTP hydrolysis. 
\PNrevA{}{We also see many \emph{correct} complexes bind and get rejected, before or after GTP hydrolysis. This is the price paid for accuracy in the classic HN proofreading model.}{}
The error rate in the simulation is slightly larger than \PNrevB{1/25 = 0.04 
Michaelis--Menten kinetics prediction for the error rate given the rate constants in Table 1 is 0.45}{$1/(\phi_3)^2 = 1/5^2 = 0.04$, however. The discrepancy is expected,}{*} because the limiting value $1/(\phi_3)^2$ is only achieved in the limit as the incorporation and hydrolysis catalytic rates are sent to zero \cite{ISI:000398884800036,Hopfield:1974p15095}. \PNrevA{Supplementary Figure 1 shows the predicted error, based on Michaelis--Menten kinetic analysis \cite{Banerjee:2017en}, with variation in incorporation or hydrolysis rate. The graph shows that the error approaches 0.04 as the incorporation rate (1a) or hydrolysis rate (1b) decrease.}{}{} 

\subsection{More realistic model\label{ss:mrm}}
Much has been learned about ribosome dynamics after Hopfield's and Ninio's original insights \cite{Arodn11a,Bbaha17a}. We now know that each step in our model consists of substeps. For example, GTP hydrolysis is subdivided into GTPase activation followed by actual hydrolysis, the latter step probably depends on a rearrangement of ``monitoring bases'' in the ribosomal RNA, and so on \cite{Satpati:2014kz}.

The  model studied in \ssref{ckp} was designed to show the HN mechanism in its ``classic,'' or pure, form, and how it can enhance fidelity even without help from the effects just described. For example, we assumed that the only dependence on right versus wrong tRNA was via unbinding rates. Indeed, such dependence was later seen at the single-molecule level \cite{Blanchard:2004p7821}. But it now appears that some of the forward rates also depend on the identity of the tRNA \cite{Zaher:2010ga,ISI:000403920000010,ISI:000393403600005}, an effect sometimes called ``internal discrimination.'' In the limit that the ribosome  uses only internal discrimination (activation barrier heights of correct and incorrect tRNA binding differ and the equilibrium constants are the same), minimum error is obtained at fast catalytic rates \PNrevA{\cite{Sartori:2013fv}}{\cite{Bennett:1979tb}}{*}. This is in contrast to the HN scheme, which achieves minimum error as catalytic rate tends to zero. \PNrevA{However, a purely internal discrimination model can only discriminate at one step and is thus not as accurate as a pure HN model for similar bias factors in the tRNA binding rates.}{}{*}

In our stochastic simulation model, it is straightforward to add internal discrimination effects by altering rate constants (Table I column 5). See Appendix \ref{a:2} for discussion of the values.

Supplemental video \ref{sv5} shows that the ribosome with experimentally measured rates can be faster and more efficient than a ribosome with only classic HN proofreading. \PNrevC{}{We see both a bias for correct tRNA binding/hydrolysis and a bias for rejection of wrong tRNAs before GTP hydrolysis.}{*} Of the 26 correct tRNA binding events in this run, 25 resulted in successful incorporation. This is compared to the  fraction $24/10245=0.002$ of productive correct tRNA binding events in a typical run of the classic HN ribosome simulation. In addition, of the 30 incorrect tRNA binding events on the realistic ribosome, all 30 resulted in rejection. The error fraction of the ribosome with realistic rates is also more accurate than the classic HN ribosome. Of 10\bn000 amino acids simulated, 18 wrong amino acids were incorporated (error fraction of 0.0018), compared to \PNrevA{X wrong amino acids}{6/100}{*} for the classic HN ribosome. \PNrevC{This matches the Michaelis--Menten kinetics predicted error fraction 0.0017 for the realistic ribosome.}{This simulated error fraction of 0.0018 is consistent with the analytic prediction of 0.0017  from first-passage times % from steady-state solutions to Michaelis-Menten differential equations 
\cite{ISI:000398884800036}.}{*}

However, the simulated ribosome with in-vitro measured rates is still not as fast or accurate as the real  \emph{E.~coli} ribosome in vivo, which translates at 15--20 amino acids per second with an error rate of 1/10\bn000 \cite{Bmilo15a}. Thus, the realistic ribosome likely evolved to combine HN proofreading (quasiequilibrium, energetic proofreading) with internal discrimination (unequal forward rates) to optimize speed, efficiency, and accuracy \PNrevA{\cite{Sartori:2013fv}}{\cite{Rao:2015vy}}{}. Despite this, simulating the classic HN model of proofreading is still a valuable exercise for students. By visualizing discrimination via only a difference in unbinding rates, students see the minimal components necessary to attain high accuracy in a broad class of biological reactions. Also, the classic HN mechanism illustrates an essential part of biological proofreading which fundamentally relies on non-equilibrium physics.

There is also recent evidence pointing to \emph{two} kinetic proofreading steps, that is, two sequential, nearly irreversible steps each of which can be followed by unbinding of tRNA \cite{ISI:000388835700066,Chen:2016bx}. Our simulation could be extended to include such effects\PNrevC{}{, whereas analytic methods would quickly become intractable}{*}.
Finally, additional interesting steps arise during ``translocation,'' in which the previous tRNA, from which the nascent peptide chain has been released, and the current tRNA, now carrying that chain with an additional amino acid, are both shifted one step inside the ribosome, freeing the binding site so that the entire cycle can begin again (\fref{f4}a). Because this step is not related to accuracy, we have simplified by omitting it from our model.

\subsection{Role of thermodynamic driving force\label{ss:rtdf}}
For comparison, we return to the classic Hopfield--Ninio model, this time operating at nearly equilibrium concentrations of GTP, GDP, and $\Pi$ to demonstrate the importance of the ``one-way door''  (the GTP hydrolysis step).  Table I column 6 summarizes the rates for this undriven model (see Appendix \ref{a:3}).
% {the thermodynamic consistency condition $\kpc  \mpc  \lc = \kc  \mc  \lpc$ was used to calculate $\mc$ and $\lpc$, with $\mc=\lpc$ (\tref{t1}, column 6).}{*}

With \PNrevA{equilibrium}{these}{*} rates, the reaction still creates a chain,  because we assumed a fixed probability per time to irreversibly add an amino acid whenever the ribosome visits its activated state. But this time a typical run gave 17 errors in a chain of length 100, illustrating the significance of the thermodynamic driving force in reducing the error rate. \PNrevA{This matches the Michaelis--Menten predicted error of 0.49 given the rate constants in Table I column 6. This error rate is consistent with the fact that, for a HN model with only difference in the unbinding rates, the error can never fall below $1/\phi_3 = 1/5$ in equilibrium conditions \cite{Hopfield:1974p15095}.}{This error rate of {about 0.17} is consistent with the Michaelis-Menten predicted error of $1/\phi_3=0.20$, which is the lowest the error can be for a classic HN model in equilibrium conditions \cite{Hopfield:1974p15095}.}{*} 

Analysis of the events in the simulation showed that of the 17 wrong amino acids incorporated, 10 were through direct binding of GDP$\cdot$tRNA. Thus, for many amino acids, the first discrimination step was bypassed\PNrevC{This results in the {\color{red}0.20 [[PREV PARAGRAPH SAID 0.17]]} error rate observed in the simulation. Thus, two independent discriminations are not possible in a simulation with GTP hydrolysis at equilibrium}{, resulting in the high error rate observed  in the simulation.}{*}

\PNrevA{}{To gain more insight into the role of the irreversible GTP hydrolysis step, some students may wish to rerun the simulation with different incorporation and hydrolysis rates. For example, a simulation with $\mpc = 25$ and $\kadd = 4.14$ results in a simulation with many tRNAs flipping between GDP and GTP states, another way in which the two discrimination steps become coupled into one.}{*}

\section{Conclusion}
The models described here show fairly elementary physical principles that lie at the heart of cell biology.
Specifically,  gene expression and kinetic proofreading are two important, fundamental topics that are well within reach of undergraduates.

A module that introduces stochastic simulation need not dominate a semester course: One class week is enough for the first exposure. Indeed, the entire simulation plus visualization in supplementary computer code \ref{sc1} consists of just \emph{seven short lines} of code, and yet it creates a valuable educational experience not available in a static textbook. Moreover, the opening material is not specifically biological in character; it can serve as
a stepping stone to more complex simulations relevant for a variety of courses. 

\section*{Acknowledgments}
We are grateful to Ned Wingreen and Anatoly Kolomeisky for correspondence and to Bruce Sherwood for help with software.
This work was partially supported by the United States 
National Science Foundation under Grants PHY--1601894 and MCB--1715823. Some of the work was done at
the Aspen Center for Physics, which is supported by NSF grant PHY--1607611, and at the Physical Biology of the Cell summer school
(Marine Biological Laboratory).

\appendix
\section{Choice of illustrative parameters}
\subsection{Classic HN proofreading (\ssref{ckp})\label{a:1}}
Here we outline our choice of rate parameters in Table I, column 4. Rather than pursue these biochemical details, students can simply be told that these are interesting values.

Zaher and Green measured $\kaddc\approx4.14\,\sunit\inv$ and $\mpc\approx25\,\sunit\inv$ \cite{Zaher:2010ga}. Interestingly, with those forward rates and the classic HN model's stipulation that only the unbinding rates differ, we found in the stochastic simulation that protein synthesis had a high error rate. For example, using the parameter  values shown Table I, column 4 but with \PNrevC{these two modifications}{the two forward rates above}{*}, a typical  run gave 49 errors out of 1000 amino acids simulated. 

In fact, this breakdown of kinetic proofreading was already predicted in Hopfield's original work, which pointed out that the error rate will only approach $1/{\phi_3}^2$ if additional conditions are met:
$$\mpc\lessim\kc\text{\ and\ }\kaddc\lessim\lc.$$
This condition gives the two binding/unbinding steps, which are both  quasi-equilibrium, time to reject a wrong aa-tRNA before hydrolysis or incorporation; it was not satisfied for the parameter values just mentioned. Thus, if the wrong aminoacyl-tRNA (aa-tRNA) bound to the ribosome, there was a low probability it would unbind before either GTP hydrolysis or incorporation.

Thus, for the purpose of illustrating a pure HN model of proofreading, we modified the experimental values of $\kaddc$ and $\mpc$ to the ones shown in Table I.

We also required illustrative values of $\mc$ and $\lpc$, which were not measured experimentally by Zaher and Green. However, these values are constrained by thermodynamic consistency \cite{Bhill89b}, which requires that even if the reaction is run far from equilibrium, nevertheless the reactions must be \emph{capable} of creating an equilibrium state. When a reaction graph contains a closed loop (cycle), as ours does, this condition requires that 
\begin{equation}k'_{\mathrm{c,eq}}m'_{\mathrm{c,eq}}\ell_{\mathrm{c,eq}}=k_{\mathrm{c,eq}}m_{\mathrm{c,eq}}\ell'_{\mathrm{c,eq}}
.\qquad\mbox{equilibrium}
\label{e:Y}\end{equation}
Assuming for simplicity that substrate-dependent reactions are first order gives
\begin{equation}\ln\frac{\kpc\mpc\lc}{\kc\mc\lpc}=\ln\frac{[\CGDP]_\mathrm{eq}}{[\Pi]_\mathrm{eq}[\CGTP]_\mathrm{eq}}+\ln\frac{[\CGTP]}{[\Pi][\CGDP]}
.\label{e:X}\end{equation}
The corresponding equation for wrong tRNAs is of the same form, except with wrong tRNA rate constants used. Dividing \eref{X} for the wrong tRNA by the equation for the right tRNA gives the consistency condition that $\phi_1\phi_2\phi_3 = \phi_{-1}\phi_{-2}\phi_{-3}$. The values in Table I  column 4 satisfy this condition.

Adamcyzk and Warshel calculated the free energy change for GTP hydrolysis on free EF-Tu via a molecular dynamics simulation as $\Delta G^{\prime0}\approx-18\,\kcalunit/\moleunit$ \cite{Adamczyk:2011jl}\PNrevC{}{, which we used to calculate the first term on the right}{*}.
For the second term on the right side of \eref{X}, [C$\cdot$GTP] was taken from Zaher and Green since the simulation's rate constants were based on those experiments. In Zaher and Green's experiments, EF-Tu$\cdot$GTP complexes were incubated with 0.5$\,\uMunit$ of aa-tRNA for 15 minutes before injection into a stopped-flow instrument. The [C$\cdot$GDP] and $[\Pi]$ can then be estimated by using 
\begin{equation}
[\CGDP]=[\Pi]=[\CGTP]\ssr{ini}(1-\ex{-k\ssr{cat}t})\approx[\CGTP]\ssr{ini}k\ssr{cat}t,
\label{e:Z}\end{equation}
where $t$ is the incubation time of 15 minutes and $k\ssr{cat}$ is the rate of GTP hydrolysis on free EF-Tu$\cdot$aa-tRNA complexes.
%From the radiolabeled substrate kinetic experiments in 
Fasano et al. calculated $k\ssr{cat}\approx5.56\eem{5}\,\sunit\inv $ \cite{Fasano:1982uw}. Thus, [\CGDP] = [\Pi] = $0.025\,\uMunit$ and $[\CGTP]=0.475\,\uMunit$.

Using these non-equilibrium concentrations and the assumption that $\mc=\lpc$, \eref{X} yields the common value of $2.36\ee{-12}\,\sunit\inv$. A similar argument gives the same values for $\mw$  and  $\lpw$. For the purposes of our simulation, however, these values are essentially zero; over the limited duration of our simulation the corresponding transitions don't occur.  We replaced them all by another  small value, $0.001\,\sunit$.

As a further confirmation that the rates chosen are appropriate for a pure Hopfield scheme, we also checked the following conditions described by Hopfield \cite{Hopfield:1974p15095}:\begin{itemize}
\item $\mpc\lessim\kc$.
\item $\kadd<\lc$.
\item $\mpc\kpw/(\mpc+\kpw)>\lpw$.
\item $\mw\lessim\lw$,\ $\kadd\lessim\lw$.
\end{itemize}

\subsection{More realistic model (\ssref{mrm})\label{a:2}}
To model a ribosome as it might work in the cell, in this simulation we more closely followed the rates measured by Zaher and Green (Table I, column 5). The rate constants
$\mpc$ and $\kaddc$ were set to their experimentally measured values of 25$\,\sunit^{-1}$ and 4.14$\,\sunit^{-1}$, respectively\PNrevA{, and $\lpc$ and $\mc$ were estimated using the same analysis in section 4.4 (see supplement X.X for details)}{}{*}. The $\phi_i$ values were also modified to match experimental measurements
\cite{ISI:000398884800036,Banerjee:2017en}.
\PNrevC{}{$\phi_2$ and $\phi_3$ were not measured in Zaher and Green, so \PNrevC{}{we follow}{*} Banerjee et al., who chose $\phi_2=1$ and got $\phi_3$ by the thermodynamic consistency condition discussed in \aref{1}.}{*} Banerjee et al.\ also chose $\phi_{-2}=1$, then got $\phi_{-3}$ by imposing the consistency condition% (their eqn. 2)
: $\phi_1\phi_2\phi_3=\phi_{-1}\phi_{-2}\phi_{-3}$.

\subsection{Model with no chemical driving force  (\ssref{rtdf})\label{a:3}}
To illustrate the importance of the GTP hydrolysis step in the HN model, rate values were chosen to simulate protein synthesis without GTP, GDP, and \Pi{} concentrations out of equilibrium. The rates chosen for the equilibrium proofreading model were similar to those in the classic HN model, except that \PNrevC{$\kpc$ and $\mpc$ were set to their experimental values and }{}{*}the equilibrium consistency condition, \PNrevB{}{\eref{Y}}{*}, was used to calculate $\lpc$ and $\mc$. \PNrevC{}{In addition, the $\phi$s were kept the same as in \aref{1}, that is, satisfying Hopfield's condition that only the unbinding rates differs between wrong and right tRNAs. }{*}Using the same assumption that  $\mc=\lpc $ for simplicity yielded $\mc =\lpc = $\PNrevBfn{13.04}{0.26}$\,\sunit^{-1}$,  higher than in the classic HN model.

\section*{Supplementary online material}
\noindent\textsl{Videos:}
\begin{enumerate}
	\item\label{sv1}\texttt{\small ChenVideo1-BeadJump.mov} at \\
	% \textbf{\color{red}TEMPORARY LINK\\}
\texttt{\small\href{http://www.physics.upenn.edu/biophys/PMLS/Media/brownian/BeadJump.mov}{http://www.physics.upenn.edu/biophys/PMLS/Media/brownian/BeadJump.mov}}:\\
Video micrograph of a micrometer-scale bead in thermal motion, hopping between the minima of a symmetric double-well potential field created by an optical trap. Courtesy Adam Simon (see \cite{Simon:1992cr}). 

\item\label{sv2}\texttt{\small ChenVideo2-Flipper.mov} at 
% \textbf{\color{red}TEMPORARY LINK} 
\texttt{\small\href{https://vimeo.com/269210861}{https://vimeo.com/269210861}}:\\
Hopping in a symmetric, two-state well.
Animation generated by supplementary computer code \ref{sc1}, reminiscent of the behavior seen in video \ref{sv1}. 

\item\label{sv3}\texttt{ChenVideo3-HNslow.mp4} % formerly HN_300s__slow_080118
%\texttt{\small Video3-slow.mov}
at 
% \textbf{\color{red}TEMPORARY LINK}
\texttt{\small\href{https://vimeo.com/283759767}{https://vimeo.com/283759767}}\thinspace
:
%\texttt{\small\href{https://vimeo.com/269210868}{https://vimeo.com/269210868}}
\\
Proofreading in the classic Hopfield--Ninio model.
Animation generated by supplementary computer code \ref{sc4}, from a simulated trajectory generated by supplementary computer code \ref{sc3} with parameters discussed in \ssref{ckp}.
In this and the following videos, 
the left green number is the frame number and the right green number is the state's sequence number. Green objects represent correct transfer-RNA/amino acid complexes; red objects represent incorrect choices. Cylindrical objects represent complexes containing GTP, arriving from solution. A sphere in the center position represents a complex containing GDP. The growing chain of spheres on the right represent amino acids incorporated into the nascent polypeptide. The simulation was done assuming cellular conditions, that is, GTP far from equilibrium with \GDP{} and inorganic phosphate. \PNrevC{Only a short initial time interval is shown.}{Only a short initial time interval is shown (the first 3\% of full simulation), so that tRNA binding, unbinding, hydrolysis, and incorporation events can be seen in detail.}{*} The frame rate was chosen to be 15 frames/second, and the total duration was chosen to be $10\,000\,\sunit$ ($150\,000$ frames). The actual simulation time is $26\,692$ seconds ($41\,185$ states). The value of $t_\mathrm{min}$ (see \ssref{vsr}) was chosen to be  $(0.5)(1/15)\,\sunit = 0.033\,\sunit$. 

\item\label{sv4}
\texttt{ChenVideo4-HNfast.mp4} %formerly HN_600s__fast_080818
at 
% \textbf{\color{red}TEMPORARY LINK}
\texttt{\small\href{https://vimeo.com/284065985}{https://vimeo.com/284065985}}\thinspace
: 
%\texttt{\small\href{https://vimeo.com/269210900}{https://vimeo.com/269210900}}
\\
The same as in video \ref{sv3}, but speeded up and covering the incorporation of 25 amino acids.
The error rate for binding was deliberately taken to be unrealistically large, $1/\phi_3=0.20$, in order to generate some errors in a short simulation. 
The frame rate was chosen to be 30 frames/second, and the total duration was chosen to be  $600\sunit$ ($18\,000$ frames). The actual simulation time is $26\,692$ seconds ($41\,185$ states). The value of $t_\mathrm{min}$ (see \ssref{vsr}) was chosen to be  $(0.42)(1/30)\,\sunit = 0.014\,\sunit$. 

\item\label{sv5}\texttt{ChenVideo5-InternalDisc.mp4} %formerly banerjee_150s_080118
at  
% \textbf{\color{red}TEMPORARY LINK}
\texttt{\small\href{https://vimeo.com/283760592}{https://vimeo.com/283760592}} \thinspace:
\\
Proofreading in a model with internal discrimination.
The same as in video \ref{sv3}, but with parameters discussed in appendix \ref{a:2}. The same chain length of 25 was used and no wrong amino acids are incorporated because the realistic ribosome has an error rate much lower than 1/25 (\ssref{mrm}).
The frame rate was chosen to be 15 frames/second, and the total duration was chosen to be 150 s (2250 frames). The actual simulation time is 8.44 seconds (138 states). The value of $t_\mathrm{min}$ (see \ssref{vsr}) was chosen to be  $(0.5)(1/15)\,\sunit = 0.033\,\sunit$. 
\end{enumerate}

\bigskip\noindent\textsl{Computer codes:}  These codes are also available from
\texttt{\small\href{https://github.com/NelsonUpenn/PNelson_code}{https://github.com/NelsonUpenn/PNelson\_{}code}}\thinspace. They
 were written in Python version 3. The \<.ipynb> files run in the Jupyter Notebook environment and use the VPython 7 package \cite{Evpyt17a}. The \<.py> files may be run in any Python implementation. One way to obtain VPython and Jupyter is from the free Anaconda distribution (\texttt{\small\href{http://anaconda.com}{http://anaconda.com}}): After regular installation, issue the command
\\
\texttt{\$ conda install -c vpython vpython}\\
to your operating system's command shell. For more details see \cite{Bkind15a}.
\begin{enumerate}
	\item\label{sc1}
	\<ChenCode1-flipper.ipynb>: Simulate two-state transitions.
	\item\label{sc2} 
	\<ChenCode2-transcrip.py>: Simulate birth-death process. % formerly transcrip2rxn.py>, \<viewTransc.py
	\item\label{sc3}
	\<ChenCode3-riboProof.py>: Simulate tRNA selection in a ribosome. % formerly riboProof.py>, \<ribosim.py
	\item\label{sc4}
	\<ChenCode4-kproofBackend.ipynb>: Display the result of Code 3 as an animation. % formerly kproof\_{}backend.ipynb
\end{enumerate}
\bibliographystyle{unsrt}%pnauthordate3}
\bibliography{AJPbibdesk,AJPpapers,AJPextra}

\end{document}